\newcommand{\apj}{ApJ}%                                         % Journal abbreviations
\newcommand{\apjs}{ApJS}

\newcommand{\aap}{A{\&}A}

\newcommand{\mnras}{MNRAS}
\newcommand{\aj}{AJ}

\documentclass[
    ,final            % use final for the camera ready runs
  ]
  {aipproc}

\layoutstyle{6x9}

\begin{document}

\title{X-ray Point Sources and Radio Galaxies in Clusters: Source of Distributed Heating of the ICM?}


\author{Quyen N. Hart}{
  address={Center for Astrophysics and Space Astronomy, Department of Astrophysical and Planetary Sciences, UCB-389, University of Colorado, Boulder, CO 80309}
}

\author{John T. Stocke}{
  address={Center for Astrophysics and Space Astronomy, Department of Astrophysical and Planetary Sciences, UCB-389, University of Colorado, Boulder, CO 80309}
}

\author{Eric J. Hallman}{
  address={Center for Astrophysics and Space Astronomy, Department of Astrophysical and Planetary Sciences, UCB-389, University of Colorado, Boulder, CO 80309}
}

\begin{abstract}
In our ongoing multi-wavelength study of cluster AGN, we find $\approx$75\% of the spectroscopically 
identified cluster X-ray point sources (XPS) with L$_{0.3-8.0keV}>10^{42}$ ergs s$^{-1}$ and 
cluster radio galaxies with P$_{1.4 GHz}>3\times10^{23}$ W Hz$^{-1}$ in 11 moderate 
redshift clusters (0.2$<$z$<$0.4) are located within 500 kpc from the cluster center.  
In addition, these sources are much more centrally concentrated than luminous cluster 
red sequence (CRS) galaxies.  With the exception of one luminous X-ray source, we find 
that cluster XPSs are hosted by passive red sequence galaxies, have X-ray colors consistent 
with an AGN power-law spectrum, and have little intrinsic obscuring columns in the 
X-ray (in agreement with previous studies).  Our cluster radio sources have properties similar 
to FR1s, but are not detected in X-ray probably because their predicted X-ray emission falls 
below our sensitivity limits. Based on the observational properties of our XPS population, 
we suggest that the cluster XPSs are low-luminosity BL Lac objects, and thus are beamed 
low-power FR 1s.  Extrapolating the X-ray luminosity function of BL Lacs and the Radio 
luminosity function of FR 1s down to fainter radio and X-ray limits, we estimate that a 
large fraction, perhaps all CRSs with L$>$L$^*$ possess relativistic jets which can inject 
energy into the ICM, potentially solving the uniform heating problem in the central region 
of clusters.
\end{abstract}

\maketitle

%%%%%%%%%%%%%%%%%%%%%%%%%%%%%%%%%%%%%%%%%%%%
%% MAINMATTER
%%%%%%%%%%%%%%%%%%%%%%%%%%%%%%%%%%%%%%%%%%%%

\section{Introduction}
Observations of galaxy clusters have provided important clues about the formation
and evolution of stars, galaxies and AGN in these systems. And based upon the
observed correlation between supermassive Black Hole mass and galaxy bulge mass,
we expect that there should be numerous AGN in the
bright ellipticals in rich clusters as well.  Radio-loud AGN have been directly implicated in cluster 
ICM heating because evacuated ``bubbles'' in the diffuse X-ray emitting gas have 
been observed that are spatially coincident with non-thermal radio emission.
But what about other non-BCG radio-loud galaxies?  Is there a connection
between cluster radio galaxies and the X-ray AGN population?  What is the spatial
distribution of this combined population?

To answer these questions we present a multi-wavelength study of cluster AGN that differs from many previous 
studies (e.g. \citep{1995AJ....109..853L,2007ApJ...664..761M}) that use
X-ray or radio observations, but not both, to detect AGN in only one waveband.  
Most cluster AGN studies have either employed clusters in a flux-limited survey or 
simply selected clusters based upon availability.  However, since most flux-limited surveys 
tend to detect the most luminous clusters at higher-$z$, this selection naturally identifies more evolved 
structures at earlier times.  This mismatch can result in comparisons which can obscure any true evolution,
since more massive clusters are placed at the beginning of the evolutionary sequence
and less massive objects at the end.  We describe a unique cluster selection method below which 
avoids the difficulty just described and present a summary of our first results.

% ***************************
\section{Sample Selection}
% ***************************
Guided by hydrodynamical simulations to track the growth of massive cluster
halos ($M>10^{15} M_\ensuremath{\odot}$) from $z\approx1$ to the present epoch, we have used the
temperature of the ICM as a proxy for cluster mass and have 
selected eleven 0.2$<$z$<$0.4 clusters that are predicted to evolve into objects like the
present-day Coma Cluster (kT=8.2$\pm$0.2 keV).  
These 11 clusters are MS 0440.5+0204, Abell 963, RX J0952.8+5153, Abell 2111, MS 1455.0+2232, Abell 1758,
MS1008.1-1224, MS2137.3-2353, Abell 1995, MS1358.4+6245, and Abell 370.
Our unique method attempts to select clusters of similar mass at a given redshift and 
thus these clusters on the ``Road to Coma'' will be used for consistent evolutionary comparisons later.
At z$=$0.3, the ICM temperature of a Coma Cluster progenitor is predicted to be 7.0$\pm$2.6 keV where
the temperature spread is due partly to the modest number of realizations of Coma Cluster mass-scale 
objects in these simulations.  The details of the cosmological simulations 
(similar to \cite{1988MNRAS.235..911E} and includes preheating of the ICM \citep{2001ApJ...555..597B}) 
will be described in a future publication.

% *******************************************
\section{Multi-wavelength Data}
% *******************************************
We re-analyzed archival {\it Chandra} observations with adequate exposure times
to detect XPSs with L$_{X}$$>$10$^{42}$ ergs s$^{-1}$.
XPSs were identified with the CIAO program {\it wavdetect}. Net XPS counts (0.3-8.0 keV) were 
estimated within the 95\% encircled energy radius and converted to X-ray fluxes by assuming a power-law 
spectral index of $\Gamma$=1.7 (N$_{E}\propto$E$^{-\Gamma}$).  
We have augmented radio survey data (e.g., NVSS, FIRST)
with an analysis of new and archival VLA 20cm continuum images to detect radio galaxies with
P$_{1.4GHz}$$>$3$\times$10$^{23}$ W Hz$^{-1}$ across the entire survey.  This limit allows the detection 
of many lower radio power FR 1 sources while excluding the lower luminosity radio sources due to star formation
(P$_{1.4GHz}$$<$10$^{22.75}$ W Hz$^{-1}$, \citep{2003ApJS..146..267M}).  

Two-color images and/or photometry are publicly available from SDSS, CNOC \citep{1996ApJS..102..269Y}, and/or 
{\it ChaMP} \citep{2004ApJS..150...43G}.
We obtained new optical images (Sloan \emph{g,r,i}) of one cluster at Apache Point Observatory (APO).
For candidate cluster AGN without published redshift and/or spectra, spectroscopic observations also were obtained
for all objects with Sloan $r<20.8$ using the APO 3.5m with the Double Imaging Spectrograph (DIS-II),
which provides complete spectral coverage from 3700--9000 \AA. 
Spectra for radio galaxies and XPSs are complete to $M_r$$<$-20.8 (M$^{*}_r\approx$-20.8).
We define passive galaxy spectra as those showing only
stellar absorption lines (e.g., the 4000\AA~Ca break, Mg Ib, Na I), while active galaxy spectra have detectable emission lines
(e.g., H$\alpha$, [O~III], etc. with [O~III]5007 \AA $>$ H$\beta$).  We do not find any example of a starburst optical
spectrum (strong emission lines with [O~III]5007 \AA $\approx$ H$\beta$) among either the radio galaxies or XPSs.

\begin{figure}
  \includegraphics[height=.35\textheight]{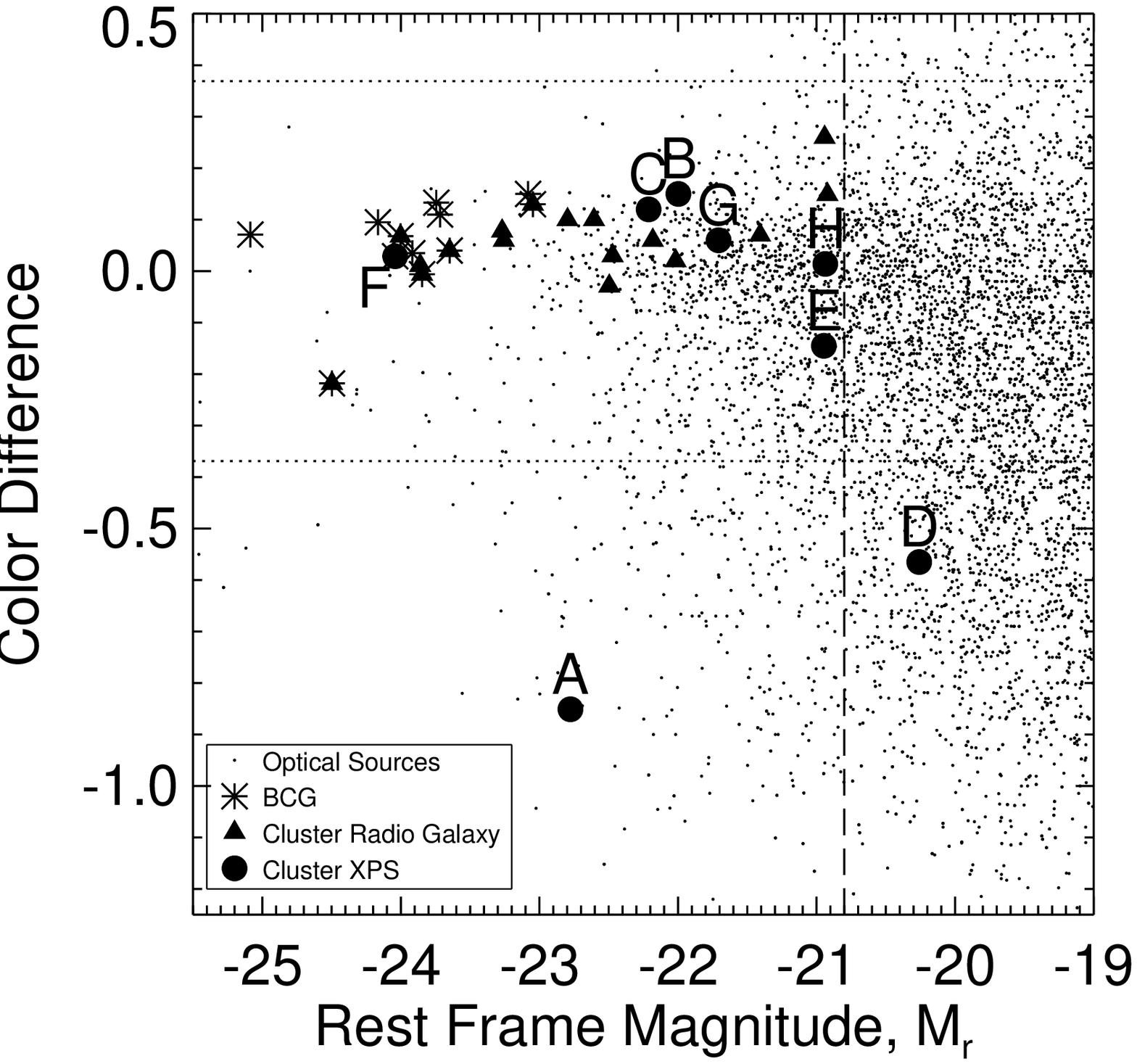}
  \includegraphics[height=.35\textheight]{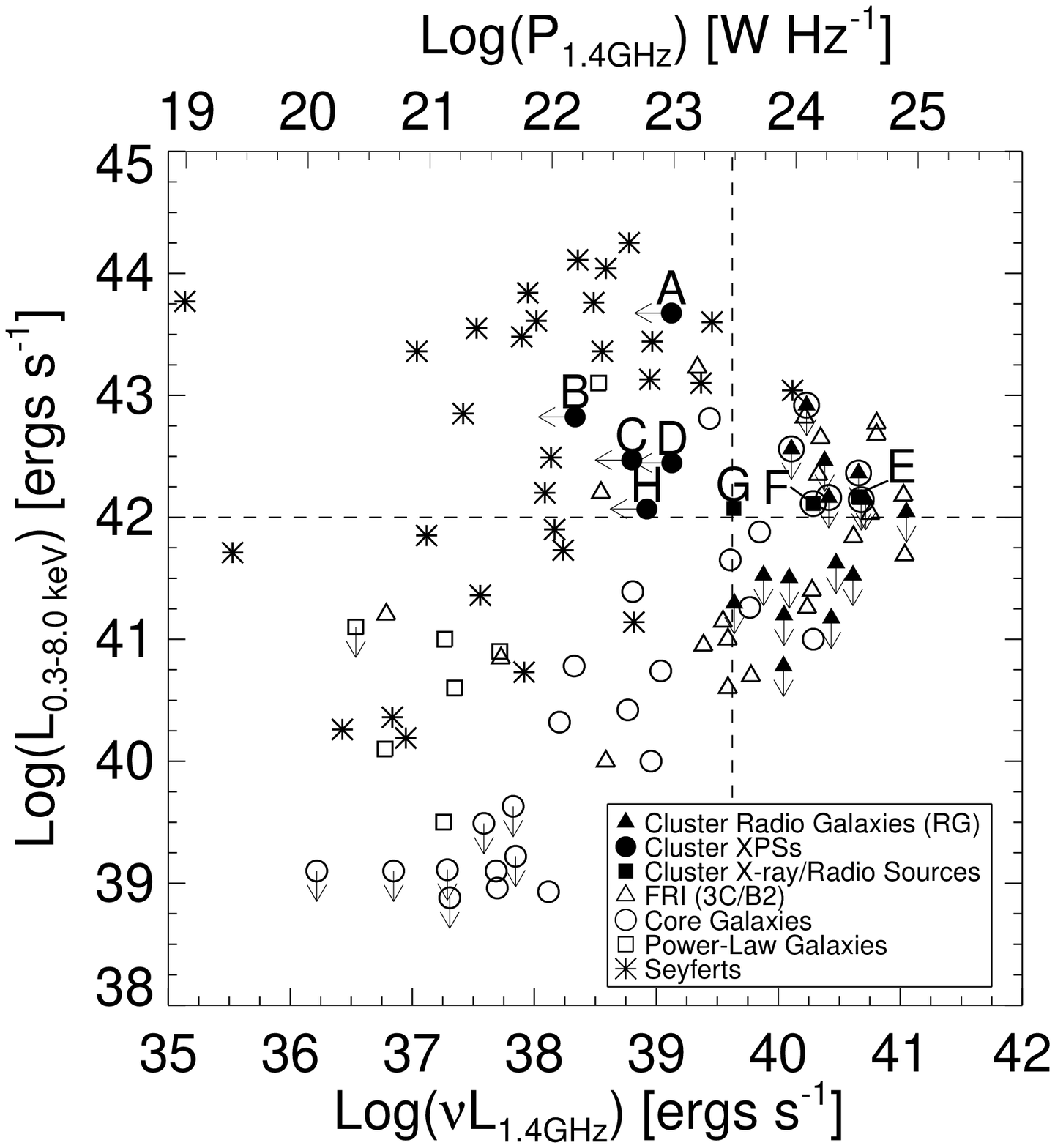}
  \caption{(\emph{Left}) Composite Color Magnitude Diagram for 11 clusters with 0.2$<$z$<$0.4:
The color difference is the difference between the mean CRS and the observed value.
The vertical dashed line marks
M$_{r}^{*}=-20.8$.  The horizontal dashed lines display the typical $3\sigma$ spread in CRS colors.  Notice that the
majority of cluster radio galaxies and XPSs lie on or near the cluster red sequence (CRS), with the exception of sources A and D.
(\emph{Right}): X-ray Luminosity versus Radio Power for cluster radio galaxies and
XPSs compared to typical FR 1 galaxies, ``core" elliptical galaxies, and Seyfert galaxies.}
\label{fig:cmd_plot}
\end{figure}

\section{Results}
Within 1 Mpc of the cluster X-ray emission centroid, we find 8 XPSs with $L_{0.3-8.0 keV}$$>$10$^{42}$ ergs~s$^{-1}$ and
20 radio galaxies with P$_{1.4GHz}$$>$3$\times$10$^{23}$ W Hz$^{-1}$, confirmed to be cluster members and have 
M$_{r}$$<$-20.8.
Figure~\ref{fig:cmd_plot} (\emph{left}) displays the composite color magnitude diagram for our 11 clusters.
%The color difference is the difference between the mean CRS and the observed value.
The majority of cluster radio galaxies lie on the CRS, as expected since they are known to reside in 
passive galaxies \citep{2003ApJS..146..267M}.
The letter identifiers (A--H) refer to our eight cluster XPSs in decreasing X-ray luminosity order.  Two cluster XPSs 
(sources A and D) are much bluer than the CRS.  Source A has an optical spectrum consistent with a Seyfert nucleus 
(with emission-line luminosity of [OIII]$>>$H$\beta$).  On the otherhand, Source D
is several tenths of a magnitude bluer than the red sequence, but, nonetheless, possesses a 
passive absorption line spectrum.  A composite X-ray spectrum of the 7 XPSs (not including Source A) can 
be fit with a power-law spectrum with $\Gamma=1.6\pm0.2$ with no evidence for intrinsic absorption and are
are hosted by luminous red galaxies with no evidence for typical Seyfert-like emission signatures.

A cumulative radial distribution plot reveals that $\approx$75\% of our cluster radio galaxies and 
XPSs are located within 500 kpc of the cluster center, 
compared to 40\% of the CRS galaxies within the same radius.  A two-sided K-S test between the full 
X-ray $+$ radio population and the CRS galaxy population as a whole
are inconsistent with being drawn from the same parent population at $>$ 99.999\% confidence level (K-S D-statistic = 0.36 and
probability = 3.3$\times$10$^{-6}$).  Within 500 kpc 16 radio galaxies and 4 XPSs make up 8$\pm$2\% of the bright 
($L>L^*\approx$ 265) CRS galaxies.  Our result strongly suggests that the triggering of a radio-loud
AGN or an XPS in these cluster galaxies is due to some, as yet undetermined, mechanism related to the ICM density such
as in the Bondi accretion model of \citet{2006MNRAS.372...21A}.  Additionally, these non-central cluster AGN sources could
provide a naturally distributed source of heating of the ICM.

%Are we really seeing two populations of sources or is this just a function of our flux limits?
Figure~\ref{fig:cmd_plot} (\emph{right}) displays L$_X$ versus P$_{1.4GHz}$ for our cluster radio galaxies and XPS, as well
as for typical FR 1s \citep{2004ApJ...617..915D}, ``core" ellipticals \citep{2006A&A...447...97B}, and Seyfert galaxies \citep{2007A&A...469...75C}.  
With the exception of three sources our cluster radio galaxies and XPSs rarely overlap with our luminosity limit.  
Within our survey limits our cluster radio galaxies are consistent with the X-ray/radio luminosities of FR 1s.  
However, if our cluster XPSs were also FR 1s, we would have detected them in our radio survey.  The XPSs appear to have
X-ray luminosities and radio power limits consistent with Seyferts, except that the XPSs have no evidence for
obscuration in the optical nor in X-rays.  However, Source D has an optical color that is $\sim$ 0.5 mag bluer 
in (g-r) than its associated CRS suggesting a possible power-law AGN excess.  We suggest that the properties of 
our cluster XPSs are very similar to low luminosity BL Lac Objects of the class now called ``High-energy-Peak'' 
BL Lacs (or HBLs) (see \citep{1999ApJ...516..145R}).  

Within 500 kpc we use our detected number of cluster radio galaxies with 
P$_{1.4GHz}$$>$3$\times$10$^{23}$ W Hz$^{-1}$ and CRS galaxies with L$>$L$^*$ 
to extrapolate the RLF of radio galaxies \citep{2000A&A...360..463M} to  
P$_{1.4GHz}$$>$10$^{21.5}$ W Hz$^{-1}$.  We estimate that $\approx$85\% of all L>L* CRS galaxies are radio-loud.
Similarly, using our cluster XPS numbers, we extrapolate the XLF of HBLs 
\citep{2000AJ....120.1626R} to L$_x>$10$^{40}$ ergs s$^{-1}$ to find 
that all virtually all L$>$L$^*$ CRS galaxies host an XPS.  Thus, we conclude that both the XPSs and the radio sources are most 
readily identified as radio-loud AGN with jets which can transfer heat into the surrounding ICM.  

%%%%%%%%%%%%%%%%%%%%%%%%%%%%%%%%%%%%%%%%%%%%%%%
% BACKMATTER
%%%%%%%%%%%%%%%%%%%%%%%%%%%%%%%%%%%%%%%%%%%%%%%%

\begin{theacknowledgments}
QNH acknowledges the support from the {\it Chandra} Theory/Archive Grant AR8-9013X.
EJH acknowledges the support from NSF AAPF AST-0702923. 
\end{theacknowledgments}




%\bibliography{cluster_agn_bibtex}

%%
%% End of file `template-6s.tex'.

\end{document}